\documentstyle[aps,multicol,prb,epsf]{revtex}
\begin{document}
\draft
\title{Quantized adiabatic quantum pumping due to interference}
\author{O. Entin-Wohlman and Amnon Aharony}
\address{School of Physics and Astronomy, Raymond and Beverly Sackler
Faculty of Exact Sciences, \\ Tel Aviv University, Tel Aviv 69978,
Israel\\} \author{ Vyacheslavs Kashcheyevs}
\address{
Institute of Solid State Physics, University of Latvia, 8
Kengaraga St., LV-1063 Riga, Latvia.}

\date{\today}
\maketitle

\begin{abstract}

Recent theoretical calculations, demonstrating that quantized
charge transfer due to adiabatically modulated potentials in
mesoscopic devices can result purely from the interference of the
electron wave functions (without invoking electron-electron
interactions) are reviewed: (1) A new formula is derived for the
pumped charge $Q$ (per period); It reproduces the Brouwer formula
without a bias, and also yields the effect of the modulating
potential on the Landauer formula in the presence of a bias. (2)
For a turnstile geometry, with time-dependent gate voltages
$V_L(t)$ and $V_R(t)$, the magnitude and sign of $Q$ are
determined by the relative position and orientation of the closed
contour traversed by the system in the $\{V_L-V_R\}$ plane,
relative to the transmission resonances in that plane. Integer
values of $Q$ (in units of $e$) are achieved when a transmission
peak falls inside the contour, and are given by the winding number
of the contour. (3) When the modulating potential is due to
surface acoustic waves, $Q$ exhibits a staircase structure, with
integer values, reminiscent of experimental observations.

\end{abstract}

\begin{multicols}{2}

\section{Introduction}

When parameters of the Hamiltonian are slowly varied as function
of time, adiabatic control of the electronic states becomes
possible. When the change is carried out periodically, such that
the Hamiltonian returns to itself after each cycle, one may pump
an integral number, ${\cal N}$, of electrons through the system
during each cycle. This results in a direct current (dc), flowing
in response to an ac signal, which, when averaged over a cycle, is
proportional to the modulation frequency times ${\cal N}e$ ($e$
being the electronic charge).\cite{altshuler} Such adiabatic
control of system parameters may be realized in nanostructures,
where either the shape-forming potential, or the tunnel couplings
between the structure and the leads (connected to the electronic
reservoirs) can be modulated in a controllable way.

The possibility to transmit an integral number of electrons per
cycle through an unbiased system, has been realized in two
systems: 1. Quantum-dot devices, connected to leads via point
contacts having small conductances which were modulated in time
(turnstile devices);\cite{kouw,switkes} 2.
Surface-acoustic-wave-based devices.\cite{talyanski} In both
examples, the observed quantization has been attributed to the
Coulomb blockade, which quantizes the number of electrons on the
device. However, the possibility to achieve quantization in `open'
nanostructures is much more intriguing. In such geometries, the
electrons are not confined to a certain region but are rather
spread over the entire device. Then, Coulomb-blockade effects are
expected to play a minor role; The question hence arises whether
quantum interference of the electronic wave function suffices to
produce quantization (as originally proposed by
Thouless\cite{altshuler}).

Here we explore the conditions for the charge pumped during a
period, $Q$, to be (almost) quantized, due to interference effects
alone, without invoking electron-electron interactions. Our
discussion begins (in Sec. 2) with the derivation of the
expression for $Q$, which serves to put limitations on the
validity of the widely-used adiabatic approximation. We continue
(in Sec. 3) with an analysis of the turnstile geometry. That
discussion points out to the connection between the conditions for
resonance transmission and integral values of $Q$. In Sec. 4 we
discuss the pump based on the surface acoustic waves (SAW's).
Finally, Sec. 5 includes concluding remarks.

\section{Transport through a periodically-modulated system}

Consider a ballistic nanostructure of arbitrary geometry,
connected by leads (denoted by $\alpha$) to electronic reservoirs
having the chemical potentials $\mu_{\alpha}$, and subject to a
potential modulated periodically in time. The current flowing
through this system will be obtained by first finding the
time-dependent scattering states, and then using them to obtain
the current.

The required time-dependent scattering solutions are derived from
a systematic expansion in the temporal derivatives of the
instantaneous solutions (that is, the scattering solutions of the
Hamiltonian in which time is `frozen'). The first-order of this
expansion yields the `adiabatic approximation'.\cite{avron} This
expansion procedure necessitates that the characteristic inverse
time-constant, $1/\tau$, which describes the time dependence of
the modulating potential, will be smaller than any characteristic
energy scale of the electrons. However, it turns out that the
expansion also requires that the amplitude of the modulating
potential will be small.

Let the system be described by the Hamiltonian
\begin{eqnarray}
{\cal H}({\bf r},t)={\cal H}_{0}({\bf r})+V({\bf r},t),
\end{eqnarray}
where the potential $V({\bf r},t)$ is assumed to be confined in
space, so that asymptotic behaviors of the scattering solutions
can be clearly defined . The Hamiltonian ${\cal H}_{0}$ consists
of the kinetic energy.  As in the usual scattering treatment, we
seek for the scattering state $\Psi_{\alpha n}$, which is excited
by the free wave $w_{\alpha n}^{-}$ (incoming in the transverse
mode $n$ of lead $\alpha$ with energy $E$), which is normalized to
carry a unit flux,
\begin{eqnarray}
&\Psi_{\alpha n}({\bf r},t)=
e^{-iEt}\Bigl (w^{-}_{\alpha n}({\bf
r})+\tilde{\chi}_{\alpha n}({\bf r},t)\Bigr ). \label{scatsol}
\end{eqnarray}
By inserting this form into the time-dependent Schr\"{o}dinger
equation, (noting that $w_{\alpha n}^{-}$ is a solution of ${\cal
H}_{0}$), $\tilde{\chi}_{\alpha n}$ can be written in terms of the
instantaneous Green function, $G^{t}(E)$,
\begin{eqnarray}
\Bigl (E-{\cal H}({\bf r},t)\Bigr )G^{t}(E;{\bf r},{\bf
r}')=\delta ({\bf r}'-{\bf r}),\label{green}
\end{eqnarray}
as follows
\begin{eqnarray} \Bigl
(G^{t}\Bigr )^{-1}\tilde{\chi}_{\alpha n}({\bf r},t)=V({\bf
r},t)w^{-}_{\alpha n}({\bf r})-i\frac{\partial\tilde{\chi}_{\alpha
n}({\bf r},t)}{\partial t}.\label{chitild}
\end{eqnarray}
Note that the time dependence of the scattered wave function,
$\tilde{\chi}_{\alpha n}({\bf r},t)$, has the same characteristic
time scale as $V$: e.g., when the modulating potential is
oscillating in time with frequency $\omega$, $\tilde \chi$
contains all harmonics.

Equation (\ref{chitild}) is solved iteratively: the temporal
derivative appearing on the right-hand-side is regarded as a small
correction. The zero-order, $\chi^{t}_{\alpha n}$, is the
scattering solution of the instantaneous Hamiltonian (in which
time appears as a parameter),
\begin{eqnarray}
\chi_{\alpha n}^{t}({\bf r})=w^{-}_{\alpha n}({\bf r})+\int d{\bf
r}'G^{t}(E;{\bf r},{\bf r}')V({\bf r}',t)w^{-}_{\alpha n}({\bf
r}').\label{chit}
\end{eqnarray}
Then the scattering solution read
\begin{eqnarray}
\chi({\bf r},t)=\chi^{t}({\bf r}) +\chi^{(1)} ({\bf
r},t)+\chi^{(2)} ({\bf r},t)+...,\label{expansion}
\end{eqnarray}
with the first-order
\begin{eqnarray}
\chi^{(1)} ({\bf r},t)=-i\int d{\bf r}'G^{t}(E;{\bf r},{\bf
r}')\dot{\chi}^{t}({\bf r}'),\label{first}
\end{eqnarray}
and the second-order
\begin{eqnarray}
\chi^{(2)} ({\bf r},t)=-i\int d{\bf r}'G^{t}(E;{\bf r},{\bf
r}')\Delta\dot{\chi}^{t}({\bf r}'),\label{second}
\end{eqnarray}
where
\begin{eqnarray}
\Delta\dot{\chi}^{t}({\bf r}')=-i\int d{\bf r}''\frac{d}{dt}\Bigl
(G^{t}(E;{\bf r}',{\bf r}'')\dot{\chi}^{t}({\bf r}'')\Bigr ).
\end{eqnarray}
and $\dot{\chi}_{\alpha n}^{t}\equiv  d\chi_{\alpha n}^{t}/dt$.
Hence, in our iterative solution, the time-dependent scattering
states are given entirely in terms of the {\it instantaneous}
solutions of the problem at hand.

In the scattering formalism \cite{Lev98,L00,OEW01} the thermal
average of the current density operator is given by
\begin{equation}
\langle {\bf j}({\bf r},t)\rangle
=\frac{e}{m}{\Im}\int\frac{dE}{2\pi}\sum_{\alpha n
}f_{\alpha}(E)\chi^{\ast}_{\alpha n }({\bf
r},t)\frac{\partial\chi_{\alpha n}({\bf r},t)}{\partial {\bf r}},
\end{equation}
where $f_{\alpha}(E)$ is the Fermi distribution in the reservoir
connected to the $\alpha$ lead. Evaluating $\langle {\bf j}({\bf
r},t)\rangle$ as ${\bf r}$ approaches $\infty$ in lead $\beta$,
and then integrating over the cross-section of that lead yields
the current $I_{\beta}(t)$ flowing into lead $\beta$ \cite{EAY}.
In the following, we analyze this current, confining for
simplicity the discussion to a nanostructure connected to left
($\ell$) and right ($r$) leads.

\subsection{The adiabatic approximation}

Consider first the net current passing through the system
utilizing the adiabatic approximation, that is, keeping only the
term (\ref{first}). Then the current flowing during a single
period of the modulating potential is\cite{EAY}
\begin{eqnarray}
I=\oint \frac{dt}{\tau}\Bigl (I_{\ell}(t)-I_{r}(t)\Bigr )=I_{\rm
bias}+I_{\rm pump}.\label{Itot}
\end{eqnarray}
The first part, $I_{\rm bias}$, flows only when the system is
biased,
\begin{eqnarray}
&&I_{\rm bias}=e\oint\frac{dt}{\tau}\int\frac{dE}{2\pi}\Bigl
(f_{\ell}(E)-f_{r}(E)\Bigr )\nonumber\\
&&\times\sum_{nm}\Biggl [2|S^{t}_{rm,\ell n}|^{2} +\Re\Bigl
(S^{t}_{\ell m,\ell n}U^{\ast}_{\ell m,\ell
n}-S^{t}_{\ell m,rn}U^{\ast}_{\ell m,rn}\nonumber\\
&&-S^{t}_{rm,\ell n}U^{\ast}_{rm,\ell
n}+S^{t}_{rm,rn}U^{\ast}_{rm,rn}\Bigr )\Biggr ],\label{ibias}
\end{eqnarray}
where $ U_{\beta m,\alpha n}=\int d{\bf r}\chi^{t}_{\beta  m}({\bf
r}) \dot{\chi}^{t}_{\alpha n}({\bf r}),$ and $S^{t}_{\beta
m,\alpha n}$ is the matrix element of the instantaneous scattering
matrix. Equation (\ref{ibias}) can be considered as a
generalization of the Landauer formula, extended to include the
effect of a time-dependent potential in the adiabatic
approximation. The second part of the current, $I_{\rm pump}$, is
established by the time-dependent potential (though it is affected
by the chemical potential difference, when the latter is applied).
Explicitly,
\begin{eqnarray}
&I_{\rm pump}=e\oint\frac{dt}{\tau}\int\frac{dE}{2\pi}
\frac{\partial (f_{\ell}(E)+f_{r}(E))}{\partial E}\nonumber\\
&\times\frac{1}{2}\sum_{m}\Bigl [\langle\chi^{t}_{\ell
m}|\dot{V}|\chi^{t}_{\ell m}\rangle -\langle\chi^{t}_{r
m}|\dot{V}|\chi^{t}_{r m}\rangle\Bigr ]. \label{ipump}
\end{eqnarray}
It can be shown\cite{EAY} that the terms in the square brackets of
(\ref{ipump}) reproduce the Brouwer\cite{brouwer} formula, derived
for an unbiased system (in which $I_{\rm pump}$ is given in terms
of temporal derivatives of the instantaneous scattering matrix).

\subsection{Corrections to the adiabatic approximation}

When the second-order in the expansion (\ref{expansion}) is
retained, one obtains the first correction to the widely-used
adiabatic approximation. We will discuss here the pumping current
beyond the adiabatic approximation for an unbiased system
connected to two single-channel leads. In that case, the current
entering lead $\beta$, $\tilde{I}_{\beta}$, is\cite{EAY}
\begin{eqnarray}
&\tilde{I}_{\beta}(t)=\frac{e}{2\pi}\int dE\Bigl (\frac{\partial
f(E)}{\partial E}\Bigr )\Bigl
[\langle\chi^{t}_{\beta}|\dot{V}|\chi^{t}_{\beta}\rangle\nonumber\\
&+\Im\Bigl (\langle\chi^{t}_{\beta}|2
\dot{V}(t)\dot{G}^{t}(E)+\ddot{V}(t)G^{t}(E)|\chi^{t}_{\beta}\rangle
\Bigr )\Bigr  ].\label{full}
\end{eqnarray}
The first term in the square brackets
is the adiabatic-approximation result. The second arises form the
second-order correction (\ref{second}) to the scattering state.

The relative magnitude of the correction compared to the
leading-order term may be accessed by noting that\cite{EAY}
\begin{eqnarray}
&\langle\chi^{t}_{\beta}|2
\dot{V}(t)\dot{G}^{t}(E)+\ddot{V}(t)G^{t}(E)|\chi^{t}_{\beta}\rangle
\nonumber\\
&=-\langle\chi^{t}_{\beta}|2\dot{V}G^{t}(E)\dot{V}+\ddot{V}|
\frac{\partial\chi^{t}_{\beta}}{\partial
E}\rangle .
\end{eqnarray}
Hence, the validity of the adiabatic approximation is not only
restricted by the smallness of $1/\tau$ dominating the temporal
derivatives. It depends as well on the energy derivatives of the
scattering states (i.e., the energy scale of the instantaneous
reflection and transmission amplitudes) and the strength of the
modulating potential itself: For the adiabatic approximation to be
valid, one should have $V\tau\ll 1$.\cite{wagner} Below, we
evaluate the pumping current in the adiabatic approximation,
keeping the above restrictions in mind.

\section{Interference effects and quantized pumping in a modulated turnstile device}

Here we investigate the turnstile pump, and in particular focus on
the correlation between resonant transmission and the magnitude of
adiabatically pumped charge. This correlation has been pointed out
by Levinson {\it et al.}\cite{YEW} and by Wei {\it et
al.}\cite{wei} The idea may be summarized generically as
follows.\cite{EA} Consider a quantum dot, connected to its
external leads by two point contacts, whose conductances are
controlled by split-gate voltages which are modulated periodically
in time. During each cycle the system follows a closed curve, the
`pumping contour', in the parameter plane spanned by the point
contact conductances. As the system parameters are varied (for
example, the gate voltage on the dot) the pumping contour distorts
and shifts, forming a Lissajous curve in the parameter plane. The
pumped charge will be (almost) quantized when the pumping contour
encircles transmission peak(s) (that is, resonances) of the
quantum dot in that parameter plane. Its magnitude (in units of
the electronic charge, $e$) and sign are determined by the winding
number of the pumping contour.

This connection between resonant transmission and pumping may be
explored by studying a simple model. Employing the tight-binding
description, we imagine the quantum dot to be coupled to
semi-infinite 1D (single-channel) leads by matrix elements
$J_{\ell}$ and $J_{r}$. Those are oscillating in time with
frequency $\omega$, such that the modulation amplitude is P and
the phase shift between the $J_{\ell}$ modulation and that of
$J_{r}$ is $2\phi$,
\begin{eqnarray}
J_{\ell}&=J_{\rm L}+{\rm P}\cos (\omega t +\phi ),
\nonumber\\
J_{\rm r}&=J_{\rm L}+{\rm P}\cos (\omega t -\phi ) .\label{jljr}
\end{eqnarray}
The point contact conductances are then given (in dimensionless
units) by X$_{\ell}\equiv J_{\ell}^{2}$ and X$_{r}\equiv
J_{r}^{2}$. As the couplings of the quantum dot to the leads are
modulated in time, $J_{\ell}$ and $J_{r}$ can attain both negative
and positive values. This reflects a modulation of the potential
shaping the dot: The tight-binding parameters $J_\ell$ and $J_{\rm
r}$, which are derived as integrals over the site `atomic' wave
functions and the oscillating potential, can have both signs. The
extreme modulation arises for $J_{\rm L}=0$, when the hopping
matrix elements which couple the dot to the leads oscillate in the
range $\{$-P,P$\}$. The conductances of the point contacts are
then modulated in the range $\{$0,P$^{2}\}$. The corresponding
Lissajous curve of the pumping is then a simple closed curve.
Another possibility is to keep the couplings finite at any time,
i.e., $J_{\rm L}\neq 0$, and to modulate the couplings around this
value. In that case the Lissajous curve may fold on itself
[compare Figs. 2 and 3 below]. This yields a rather rich behavior
of the pumped charge.

The quantum dot is modeled by a `bunch' of tight-binding sites
connected among themselves. For simplicity, we take the latter in
the form of a finite chain of $N$ sites, each having the on-site
energy $\epsilon_{0}$, and attached to its nearest-neighbor with a
transfer amplitude $-J_{\rm D}$. This structure is connected to
electronic reservoirs by two 1D chains of sites, whose on-site
energies vanish, and whose nearest-neighbor transfer amplitudes
are denoted by $-J$. The Fermi energy of an electron of wave
vector $k$ moving on the leads is
\begin{equation}
E_{k}=-2J\cos ka,\label{fermi}
\end{equation}
where $a$ is the lattice constant.

\begin{figure}[h]
\leavevmode \epsfclipon \epsfxsize=7truecm
\vbox{\epsfbox{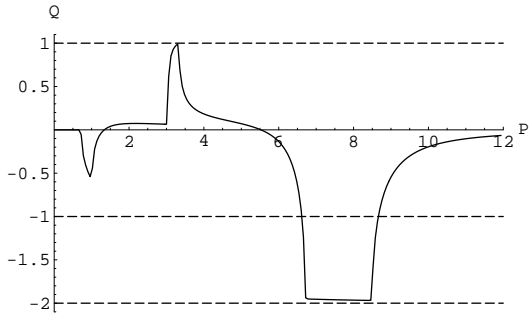}}
\caption{The pumped charge, Q, in units of $e$, as function of the
modulating amplitude, P. The parameters used are:
$\epsilon_{0}=0$, $J_{\rm D}=1$, $J_{\rm L}=2$, $ka =0.001\pi$,
$N=4$, and $\phi =0.05$. }\label{fig1}
\end{figure}

In the adiabatic approximation, the charge, Q, pumped through the
quantum dot during a single period of the modulation is given by
(\ref{ipump}) as follows
\begin{eqnarray}
{\rm Q}=\frac{e}{4\pi}\oint dt\Bigl
[\langle\chi^{t}_{r}|\dot{V}|\chi^{t}_{r}\rangle
-\langle\chi^{t}_{\ell}|\dot{V}|\chi^{t}_{\ell}\rangle\Bigr ].
\label{charge}
\end{eqnarray}
(The limit of zero temperature is taken for simplicity; Obviously
finite temperature will `smear' the effect.\cite{YEW}) The
temporal derivative of the modulating potential in the present
case is
\begin{eqnarray}
\dot{V}(n,n')=&-\dot{J}_{\ell}\Bigl (\delta_{n,1}\delta_{n',0}
+\delta_{n',1}\delta_{n,0}\Bigr )\nonumber\\
&-\dot{J}_{r}\Bigl
(\delta_{n,N}\delta_{n',N+1}+\delta_{n',N}\delta_{n,N+1}\Bigr ).
\end{eqnarray}
The expression (\ref{charge}) requires the knowledge of the
instantaneous scattering states. These are easily derived. One
writes those states in terms of the instantaneous reflection
($r_{t}$ and $r'_{t}$) and transmission ($t_{t}$) amplitudes: For
the scattering state incited by a free wave incoming from the left
one has $ \chi^{t}_{\ell}(x)=A_{0,\ell} [e^{ikx}+r_{t}e^{-ikx}]$
on the left lead, and $\chi^{t}_{\ell}(x)=A_{0,\ell}t_{t}e^{ikx}$
on the right lead, with $x=na$. Similarly,
$\chi^{t}_{r}(x)=A_{0,r} [e^{-ikx} +r'_{t}e^{ikx}]$ on the right
lead, and $\chi^{t}_{r}(x)=A_{0,r}t_{t}e^{-ikx}$ on the left lead,
for the scattering state incited by a plane wave incoming from the
right reservoir. For both, the normalization to a unit flux
implies that $A_{0,\ell}=A_{0,r}=(2J\sin ka)^{-1/2}$. The
reflection and transmission amplitudes are given by (all energies
are scaled in units of $J$)
\begin{eqnarray}
&r_{t}e^{-i2ka}+1=\Bigl (e^{ika}{\rm X}_{\ell}{\rm X}_{r}\sin
N_{-}qa -J_{\rm D}{\rm X}_{\ell}\sin Nqa\Bigr
)M_{k},\nonumber\\
&r'_{t}e^{i2Nka}+1=\Bigl (e^{ika}{\rm X}_{\ell}{\rm X}_{r}\sin
N_{-}qa -J_{\rm D}{\rm X}_{r}\sin Nqa\Bigr
)M_{k},\nonumber\\
&t_{t}=-e^{-ikN_{-}a}J_{\ell}J_{r}J_{\rm D}\sin qa
M_{k},\label{scattering}
\end{eqnarray}
with $N_{\pm}\equiv N\pm 1$, and
\begin{eqnarray}
M_{k}=&2i\sin ka\Bigl [ J^{2}_{\rm D}\sin N_{+}qa-J_{\rm
D}e^{ika}({\rm X}_{\ell}+{\rm X}_{r})\sin Nqa
\nonumber\\
&+e^{i2ka}{\rm X}_{\ell}{\rm X}_{r}\sin N_{-}qa\Bigr ]^{-1}.
\end{eqnarray}
The wave vector $q$ describes the propagation of the wave on the
quantum dot, such that $E_{k}-\epsilon_{0}=-2J_{\rm D}\cos qa$. By
using these expressions in (\ref{charge}) one finds that, as
function of the modulating amplitude, (or alternatively the gate
voltage on the dot), the pumped charge can vary as depicted, for
example, in Fig. \ref{fig1}.

\vspace{0.8cm}

\begin{figure}[h]
\leavevmode \epsfclipon \epsfxsize=5truecm
\vbox{\epsfbox{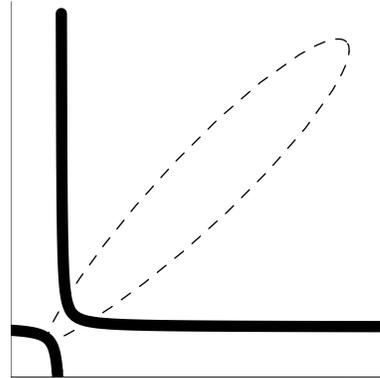}}
\caption{The pumping contour (dashed line) and the resonance lines
(thick lines), in the \{X$_{\ell}$-X$_{r}$\} plane, for P=1.
}\label{fig2}
\end{figure}

We next make the correspondence between the `quantized' values of
$Q$, as portrayed in Fig. \ref{fig1}, and the location of the
pumping contour relative to the resonant transmission of the
quantum dot. The transmission coefficient of our system,
T=$|t_{t}|^{2}$, is [see Eq. (\ref{scattering})]
\begin{equation}
{\rm T} =\Bigl [1+\frac{{\rm Z}^{2}+( J_{\rm D}\sin ka\sin
Nqa({\rm X}_{\ell}-{\rm X}_{r}))^{2}}{(2J_{\rm D}\sin ka\sin
qa)^{2}{\rm X}_{\ell}{\rm X}_{r}}\Bigr ]^{-1},
\end{equation}
with ${\rm Z}=J_{\rm D}^{2}\sin N_{+}qa +\frac{E_{k}}{2}J_{\rm
D}\sin Nqa ({\rm X}_{\ell}+{\rm X}_{r}) +{\rm X}_{\ell}{\rm
X}_{r}\sin N_{-}qa$ . Clearly, one has T=1 when X$_\ell={\rm X}_r$
and Z=0. For $N>1$, these equations give two points on the
diagonal in the \{X$_\ell-$X$_r\}$ plane. The maxima of T, when
either X$_\ell$ or X$_r$ is varied while the other parameter is
kept fixed, occur on two `resonance lines', shown in thick lines
in Figs. 2 and 3. The figures also show the topology of the curve
traversed by the system during the pumping cycle for
representative values of P. In Fig. \ref{fig2} we have P=1. The
pumping contour encloses a small part of the upper resonance line,
and also touches the peak on the lower resonance line. Indeed, Q
has an intermediate value near $-0.5$, decreasing to zero as P
moves away from 1 (see Fig. \ref{fig1}).

\vspace{0.3cm}
Increasing the amplitude to the value P=5 reveals that the pumping
contour encloses both peaks on the resonance lines, and therefore
their separate contributions almost cancel one another, leading to
a tiny value of Q. Also, the pumping curve begins to fold on
itself, giving rise to the `bubble' close to the origin (see Fig.
3). Following the increase of that bubble as P is enhanced leads
to the situation in which the bubble encloses the lower resonance,
and then the charge attains a unit value (see Fig. \ref{fig1}). As
the bubble increases further, capturing the two resonance lines, Q
again becomes very small. But upon further increasing P, we reach
the interesting situation, depicted in Fig. 3 for P=8, in which
the bubble encircles {\it twice} the upper resonance line, leading
to a pumped charge very close to $|2e|$ (see Fig. \ref{fig1}).

\vspace{0.6cm}
\begin{figure}[h]
\leavevmode \epsfclipon \epsfxsize=3truecm
\centerline{\vbox{\epsfbox{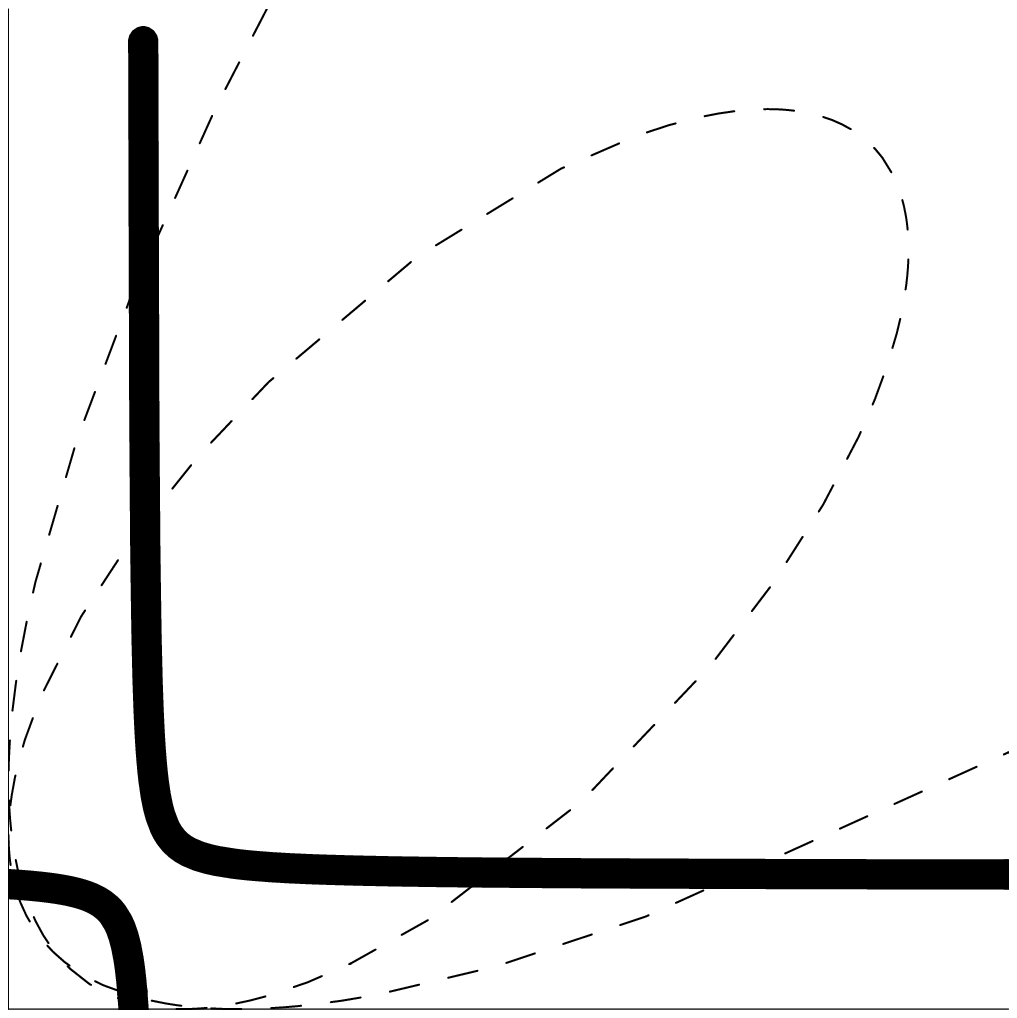}},\hspace{0.3cm},\leavevmode
\epsfclipon \epsfxsize=3truecm\vbox{\epsfbox{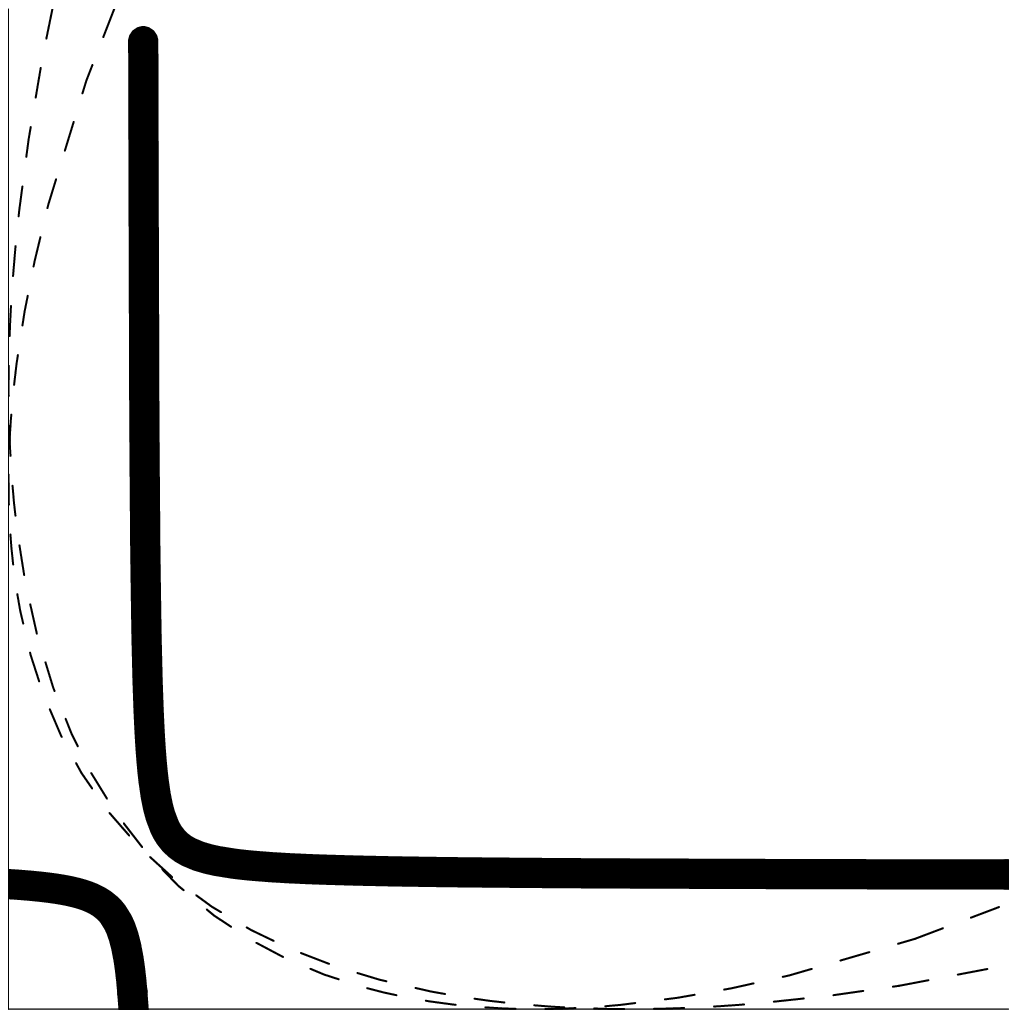}}}
\vspace{0.6cm}
\caption{Same as Fig. \ref{fig2}, for P=5 (left), P=8 (right).
}\label{fig3}
\end{figure}

Thus, the condition for obtaining integral values of the pumped
charge is that the contour traversed by the system in the
parameter plane spanned by the pumping parameters should encircle
a significant portion of a resonance line in that plane. The
magnitude and the sign of the pumped charge are determined by that
portion, and by the direction along which the resonance line is
encompassed. We emphasize that the pumping contour, as well as the
resonance lines, can be determined experimentally.\cite{YEW}

The reason for this topological description of adiabatic charge
pumping can be traced back to the expression for the pumped
charge, Eq. (\ref{charge}). The main contribution to the temporal
integration there comes from the poles of the integrand. The same
poles are also responsible for the resonant states of the
nanostructure, that is, for the maxima in the transmission
coefficient.\cite{YEW,EA}  One can imagine more complex scenarios:
Including higher harmonics of $\omega$ in the time dependence of
the point contact conductances can create more complex Lissajous
contours, which might encircle portions of the resonance lines
more times, yielding higher quantized values of the pumped charge.

\section{Interference effects and quantized pumping in a quantum
channel modulated by surface acoustic waves}

Modulation of the potential acting on a nanostructure may be also
achieved through the piezoelectric effect of surface acoustic
waves (SAW's), which is relatively large in GaAs. In the
two-dimensional electron gas formed in GaAs-AlGaAs samples the
potential created by the SAW is screened out. However, the SAW's
are effective within a quasi-one-dimensional channel (where
screening is diminished) defined in GaAs-AlGaAs samples. In the
experiments,\cite{talyanski} the time-average current exhibits
steps between plateaus, at quantized values of integer$\times
e(\omega /2\pi )$, ($\omega $ is the SAW frequency), as function
of either the gate voltage on the quantum channel, or the SAW
amplitude. Here we propose an explanation for this observation, in
terms of interference of non-interacting
electrons.\cite{SAW1,SAW2}

Unlike the turnstile-like case, the piezoelectric potential,
${\cal H}_{\rm SAW}({\bf r},t)=$P$ \cos(\omega t-{\bf q}\cdot{\bf
r})$, generated by the SAW oscillates with time everywhere inside
the nanostructure. The induced average current (in the absence of
bias), flows in the direction of the SAW wave vector, ${\bf q}$. A
realistic treatment of the experimental geometry\cite{SAW2} only
allowed a calculation at low SAW amplitude, P, yielding $Q \propto
$P$^2$. The screening of the piezoelectric potential in the wide
banks of the channel is also difficult to treat exactly. In view
of this it is useful to gain insight into the phenomenon by
applying a simple model:\cite{SAW1} A 1D channel, connected to 1D
leads. The Fermi energy of an electron moving on the leads is then
again given by (\ref{fermi}), while the `channel' Hamiltonian is
\begin{equation}
{\cal H}_{\rm osc}=\sum_n \{\epsilon_n (t)|n \rangle \langle
n|-J_n(|n \rangle \langle n+1 |+ hc) \}, \label{Hosc}
\end{equation}
with $J_n=J_D$ inside the channel,  $1 \le n \le N-1$, and
$J_0=J_\ell$, $J_N=J_r$ for the `contacts' with the leads. The
electric field generated by the SAW's
\begin{equation}
\epsilon_n(t)=V+ {\rm P} \cos[\omega t- qa(n-n_0)] \label{eps}
\end{equation}
acts only inside the channel. (Effects due to gradual screening,
or reflections from the channel ends, can be incorporated as
well.\cite{SAW1}) Here $V$ represents the gate voltage and P$>0$,
so that $\epsilon_n$ has a maximum (minimum) in the center of the
channel $n_0=(N+1)/2$ at $t=0$ ($\tau/2$).

The adiabatic approximation seems to be particularly adequate for
SAW frequencies, as $\hbar\omega$ is small compared to the
relevant electronic energy scales. We hence use (\ref{charge}),
with $\dot{V}$ there replaced by $\dot{{\cal H}}_{\rm osc}$. The
required instantaneous scattering states are straightforwardly
found; The resulting expression can be put in the form\cite{SAW1}
\begin{equation}
Q=\frac{e J_\ell^2 \sin ka}{\pi J} \int_{-\tau/2}^{\tau/2} dt
\sum_{n=1}^N \dot{\epsilon}_{n}|g_{n,1}|^{2},
\label{int}
\end{equation}
where the (time-dependent) matrix $g$ is given by
\begin{eqnarray}
\bigl (g^{-1}(E)\bigr )_{n,n'}= E \delta_{n,n'}-\bigl ({\cal
H}_{\rm osc} \bigr )_{n,n'}\nonumber\\
+\delta_{n,n'}e^{ika}\bigl
(\delta_{n,1}J_{\ell}^{2}+\delta_{n,N} J_{r}^{2}\bigr )/J.
\label{MM}
\end{eqnarray}
Using the same notations, the instantaneous transmission
coefficient of the quantum channel reads T$^t=4|g_{N,1}|^2(J_\ell
J_r/J)^2 \sin^2 ka$. This has maxima, as function of $\cos \omega
t$, at the $N$ poles of $g_{N,1}$, i.e. at the zeroes of $D(\cos
\omega t)\equiv\det  g^{-1}$. The time-averaged transmission,
$\bar{\rm T}=\int_{-\tau/2}^{\tau/2} dt{\rm  T}^t/\tau$ thus
exhibits peaks wherever such poles occur within the period $\tau
=2\pi /\omega$. Similarly, $Q$ will have singularities whenever
$\cos \omega t$ comes close to a zero of $D$ within the
integration. For small $(J_\ell^2+J_r^2)/J$, these poles have
small imaginary parts, and $Q$ exhibits large changes as $\cos
\omega t$ passes near such a pole. These steps occur exactly where
$\bar {\rm T}$ has spikes, originating from the same poles. Thus,
again, we find an intimate relation between `quantized' values of
the pumped charge, and resonant transmission.

\begin{figure}[h]
\leavevmode \epsfclipon \epsfxsize=6truecm
\vbox{\epsfbox{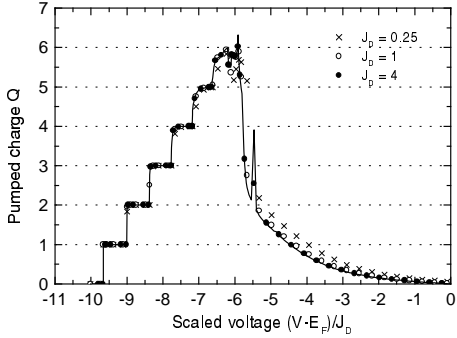}}
\vspace{1.cm} \caption{The pumped charge {\it vs.} $V$. Energies
are measured from the Fermi energy and scaled by $J_{D}$. This
renders $Q$ to depend on $J^{2}_{L}/J_{D}$ alone.}\label{fig4}
\end{figure}

Figure \ref{fig4} shows $Q/e$ {\it vs.} $V$, for $N=10$ at zero
temperature.  The other parameters used are: $J_{L}\equiv
J_{\ell}=J_{r}=0.4$, P$/J_{D}=8$, $ka=\pi /100$, and the SAW wave
length is taken to be 4 times the channel length. Several plateaus
are clearly observed, with $Q/e$ very close to an integer, ${\cal
N}=1,...,5$. (We have found that, quite generally, the number
${\cal N}$ of sharp plateaus is up to ${\cal N}=N/2$, with
possibly several additional rounded peaks or spikes.) The steps,
at $V_{\cal N}$, between these plateaus appear to be
equidistant.\cite{SAW1} For large $N$
\begin{equation}
V_{\cal N} \approx E_F \pm\bigl (P+2 J_D-\Delta({\cal
N}+\frac{1}{2})\bigr ), \label{delta}
\end{equation}
with $\Delta = qa \sqrt{2 {\rm P} J_D}$, that is, as the SAW
amplitude P increases, the steps move outwards and broaden, in
apparently accordance with experiments.\cite{talyanski}
 \vspace{0.3cm}

\begin{figure}[h]
\leavevmode \epsfclipon \epsfxsize=7truecm
\vbox{\epsfbox{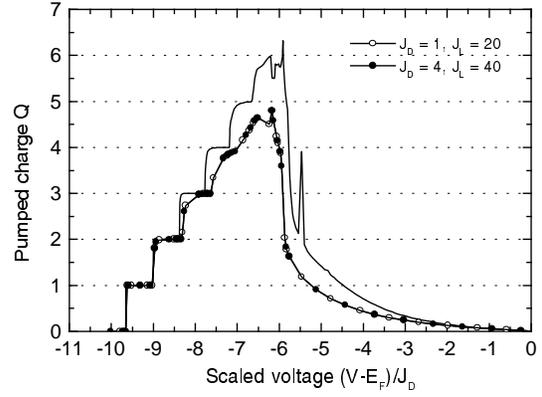}}
\vspace{1.cm} \caption{The dependence of the pumped charge on the
gate voltage, see text. }\label{fig5}
\end{figure}

Obviously, a finite pumped charge requires finite SAW amplitude P.
Indeed, the steps become rounded as P, $|E_F-V|$ or $qa$ decreases
with the rounding beginning at the larger ${\cal N}$'s; The
plateaus at ${\cal N}=\pm 1$ disappear last. These results remain
robust\cite{SAW1} over a wide range of $ka,~J_L$ and $J_D$,
provided $0<J_L^2/J \le J_D \ll {\rm P},~|E_F-V|$; As an example,
Fig. \ref{fig5} shows the effects of increasing $J_D$ and $J_L$.
As Eq. (\ref{delta}) implies, the staircase structure of $Q$ is
also obtained as function of P, at fixed $V$: $Q$ remains very
small up to P$_0=E_F-V-2 J_D+\Delta/2$, exhibits $N/2$ steps, at
intervals $\Delta$ (which now increases with $P$), and then
decreases gradually towards zero. Thus, both $V$ and P can be used
for on-off switching of the pumped current.
 \vspace{0.3cm}

\begin{figure}[h]
\leavevmode \epsfclipon \epsfxsize=7truecm\epsfysize=2.truecm
\vbox{\epsfbox{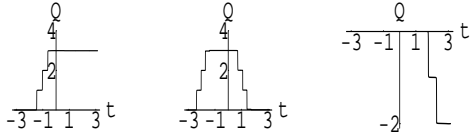}}
\vspace{1.4cm} \caption{Partial pumped charge $Q$ in units of $e$,
up to time $t$ within a period, for different values of the gate
voltage. The $t$-axis shows $\omega t$ between $-\pi$ and $\pi$.
Here P=8, $N=6$, $J_{D}=1$, $J_{L}=0.4$, $qa=\pi /10$, and $ka
=\pi/100$, and $V=-8.6$, -2, and 5.3. }\label{fig6}
\end{figure}

As mentioned above, the steps in $Q$ are correlated with the
spikes of the averaged transmission.  It is interesting to follow
these poles as function of the various parameters. Figure
\ref{fig6} depicts the partial charge $Q(t)$, resulting from
integration of (\ref{int}) only up to $t<\tau/2$, at different
values of the gate voltage. As $V$ increases through the first
step at $V_1$, $Q(t)$ suddenly exhibits a step from zero to one,
which appears at $t=0$. Upon further increasing $V$, this step
moves to the left, and at $V=V_{\cal N}$, a new step (from ${\cal
N}-1$ to ${\cal N}$) enters at $t=0$ (see the left panel of Fig.
\ref{fig6}). Afterwards, there begin to enter steps of $-1$, until
at $V=E_F$ there are exactly $N/2$ steps of $+1$ followed by $N/2$
steps of $-1$, yielding $Q=0$(the center panel of Fig.
\ref{fig6}). A similar build-up of (negative) steps occurs
starting from positive $V$-values which are then decreased, (right
panel of Fig. \ref{fig6}). Thus, unlike the Coulomb blockade
picture, in which ${\cal N}$ electrons move together, carried by a
single minimum of the moving potential, in the interference
description, $Q(t)$ changes by discrete steps of 1, implying
separate motion of the electrons, building up to ${\cal N}$ after
a full period. The step-like time dependence of $Q(t)$ implies the
appearance of higher  harmonics in $\omega$, with amplitudes
exhibiting staircase structure as well.\cite{SAW1}  Apparently,
this will not be true in Coulomb-blockade-type description.
Measurements of the induced current noise for the ${\cal N}>1$
plateaus may be useful to distinguish between the simultaneous
motion of ${\cal N}$ electrons, and the single electron
steps.\cite{talyanski}

To conclude this section, we note that even within this simple 1D
model, it is possible to study effects arising from variations of
the SAW amplitude $P$ in space (due to screening effects, or to
multiple reflections from the channel's ends), or from random
energies $\{V_n\}$, which may represent impurities within the
channel.

\vspace{1.3cm}
\vspace{1.3cm}

\section{Concluding remarks}

Using simplified models, we have demonstrated that interference
effects suffice to produce transfer of integral number of
electrons through a mesoscopic (unbiased) system, subject to a
periodically-varying potential. The calculations presented above
utilize the expression for the pumped charge, $Q$, in the
adiabatic approximation, that is, keeping only the first-order of
the expansion (\ref{expansion}). However, the next-order
corrections need not be small; It is therefore an open question
whether the quantization, and its relation to resonant
transmission, will still be present when higher terms in the
temporal expansion are retained.



\section*{Acknowledgements}

We thank Y. Imry, Y. Levinson, and P. W\"{o}lfle for helpful
conversations. This research was carried out in a center of
excellence supported by the Israel Science Foundation, and was
supported in part  by the Albert Einstein Minerva Center for
Theoretical Physics at the Weizmann Institute of Science.

\end{multicols}

\begin{references}
\bibitem{altshuler} D. J. Thouless: Phys. Rev. B {\bf 27} (1983) 6083;
B. L. Altshuler and L. I. Glazman: Science
{\bf 283} (1999) 1864.
\bibitem{kouw}
L. P. Kouwenhoven {\it et al.}:Z. Phys. {\bf B85} (1991) 381.
\bibitem{switkes} M. Switkes {\it et al.}: Science {\bf 283}
(1999) 381.
\bibitem{talyanski} J. M.
Shilton {\it et al.}
: J. Phys.: Condens. Matter {\bf 8} (1996) L531; V. I. Talyanskii
{\it et al.}
: Phys. Rev. B {\bf 56} (1997) 15180; A. M. Robinson {\it et al.}:
Phys. Rev. B {\bf 65} (2002) 045313.
\bibitem{avron} J. E. Avron {\it et al}:
Phys. Rev. Lett. {\bf 87} (2001) 236601.
\bibitem{Lev98} Y. Levinson and P. W\"{o}lfle: Phys. Rev. Lett. {\bf 83} (1999) 1399.
\bibitem{L00} Y. Levinson:
Phys. Rev. B {\bf 61} (2000) 4748.
\bibitem{OEW01} O. Entin-Wohlman, Y. Levinson, and P. W\"{o}lfle:
Phys. Rev. B {\bf 64} (2001) 195308.
\bibitem{EAY} O. Entin-Wohlman, A. Aharony, and Y. Levinson: Phys.
Rev. B {\bf 65} (2002) 195411.
\bibitem{brouwer} P. W. Brouwer: Phys. Rev. B {\bf 58} (1998)
R10135.
\bibitem{wagner} M. Wagner: Phys. Rev. A {\bf 51} (1995) 798.
\bibitem{YEW} Y. Levinson, O. Entin-Wohlman, and P.
W\"{o}lfle: Physica A {\bf 302} (2001) 335.
\bibitem{wei}
Y. Wei, J. Wang, and H. Gou: Phys. Rev. B {\bf 62} (2000) 9947; Y.
Wei, J. Wang, H. Gou, and C. Roland: Phys. Rev. B {\bf 64} (2001)
115321.
\bibitem{EA} O. Entin-Wohlman and A. Aharony: Phys. Rev. B {\bf 66} (2002) 035329.
\bibitem{SAW1}A. Aharony and O. Entin-Wohlman: Phys. Rev. B {\bf
65} (2002) 241401; V. Kashcheyevs, A. Aharony, and O.
Entin-Wohlman: unpublished.
\bibitem{SAW2} Y. Levinson, O. Entin-Wohlman, and P. W\"{o}lfle:
Phys. Rev. Lett. {\bf 85} (2000) 634.
\end{references}
\end{document}